\documentstyle[11pt,psfig]{article}
\def\et{\em et al. \em}

\def\gte{\,\lower.6ex\hbox{$\buildrel >\over \sim$} \, }
\def\lte{\,\lower.6ex\hbox{$\buildrel <\over \sim$} \, }

\begin{document}
\begin{center}
\begin{Large}
  {\bf New Determination of the Hubble Parameter Using the Principle of
Terrestrial Mediocrity}\\
\end{Large}
\end{center}

\noindent Simon\,P.\,Goodwin \\
\begin{small}
Astronomy Centre, University of Sussex, Falmer, Brighton, BN1 9QH
\end{small}

\noindent John Gribbin \\
\begin{small}
Astronomy Centre, University of Sussex, Falmer, Brighton, BN1 9QH
\end{small}

\noindent Martin\,A.\,Hendry \\
\begin{small}
Department of Physics and Astronomy, University of Glasgow, Glasgow G12 8QQ
\end{small}

\vspace{0.5cm}

\section{Abstract}

Measurements of the linear diameters of 12 nearby spiral galaxies with 
distances determined from primary indicators suggest that both the 
Milky Way Galaxy and M31 are in the middle of the range of sizes for 
such galaxies. By comparing the measured linear diameters of these 
nearby systems with the inferred diameters of a sample of more than 
3000 spirals with known redshifts, we conclude that the most likely 
value of the Hubble Parameter lies in the range
50 - 55 km s$^{-1}$ Mpc$^{-1}$.

\section{Introduction}

It is widely accepted that our Milky Way Galaxy and M31 are larger than 
most spirals (see, for example, van der Kruit, 1990).  This belief is 
largely a historical accident, resulting from early high estimates of 
the Hubble Parameter, $H_0$.  Until recently, there has been no direct way 
to test this assumption; analysis of volume limited surveys of galaxies
in the Local Group certainly reveals that the Milky Way is among the
largest galaxies in our immediate vicinity when we consider galaxies of
all Hubble types, but since the Local Group is dominated in number by
dwarf galaxies this is hardly surprising. We have found, on the other
hand, from an analysis of nearby spirals of similar Hubble type to our
galaxy, and with well-determined distances, that the Milky Way is
very much average in size (Goodwin et al., 1997).  
This is strong evidence in support of the ``principle of terrestrial 
mediocrity" (Vilenkin, 1995), that we live in an ordinary galaxy in an 
ordinary part of the Universe.

If the linear sizes of enough nearby galaxies can be determined, a 
comparison of these sizes with the inferred sizes of galaxies at higher 
redshift gives an indication of the value of $H_0$.

Even without knowing whether the sizes of the Milky Way and M31 
are typical for spirals, some researchers have attempted to use this 
kind of argument to find a value for $H_0$.  Notably, Sandage (1993a,b; 1996)
chose M101 as a ``typical" spiral, and argued that if the linear  
diameter of M101 is equal to the mean of field galaxies with similar  
appearance then the most probable value for $H_0$ is 43 
km s$^{-1}$ Mpc$^{-1}$. A key step in his chain of inference was the
assumption ``that M101 not be the largest in a distance-limited sample".  
Unfortunately, we have found that M101 is indeed one of the largest 
galaxies in our part of the Universe. A similar analysis in Sandage (1996)
using M31 as the ``typical'' spiral yielded $H_0 = 43$ km s$^{-1}$ Mpc$^{-1}$.

In this paper we adopt a similar method to estimate $H_0$, based on the
principle that a sample of nearby spirals of specific Hubble type
represents a ``fair'' sample of the intrinsic population. We extend the
range of Hubble types considered, however, in order to increase the
size of both the calibrating sample of nearby objects and the distant
population of several thousand field spiral galaxies. (Sandage's analysis used 
60 galaxies). We, therefore, aim to find in this way a more reliable value
for $H_0$. In this report we summarise our preliminary results.

\section{Galaxy diameters}

The corrected 25 B-mag arcsec$^{-2}$ isophotal diameters for all of our 
galaxies were taken from the RC3 catalogue (de Vaucoulers \et 1991).  The 
observed major axis of this isophote was corrected for 
galactic extinction but not for inclination as spiral galaxies are found 
by those authors to be ``substantially optically thick" at this isophotal 
level.  This provides us with a uniform sample using a surface brightness 
diameter that is hence independent of distance (the redshifts involved 
are too low to require any $k$ corrections) allowing us to directly compare 
our local sample with the more distant sample.

\subsection{Local calibrators}

A sample of 11 spiral galaxies with independent Cepheid distances was 
chosen to act as the local calibrators. The Cepheid distances have been 
tabulated in Giovanelli (1996) and Freedman (1996); the figure for 
NGC3351 comes from Graham \et (1997).  Table 1 
summarises the data, converting the corrected isophotal diameters in 
arcmin to a linear diameter in kpc using the Cepheid distances. 

In our earlier paper (Goodwin \et 1997) we also make a 
calculation for the Galactic diameter at this isophote using data from 
van der Kruit (1987, 1990) which yields a 25 B-mag arsec$^{-2}$
diameter of 

\[
d_{\rm 25(true)} = 26.8 \pm 1.1  {\rm kpc}
\]

leading to the conclusion, from comparison with the data in Table 1, 
that the Milky Way is a very average-sized galaxy, especially for its 
Hubble type, $2 < T < 6$ (van der Kruit 1987).  This small sample of  
galaxies also seems approximately to confirm the idea that linear diameter  
is relatively independent of Hubble type within this range (de Jong 1996). 

\begin{center}
\begin{table}
\begin{center}
\begin{tabular}{|c|c|c|c|c|c|} \hline
NGC & M & T & $d_{25 {\rm (ang)}}$ & $R$ & $D_{25 {\rm (true)}}$ \\
 & & & arcmin & Mpc & kpc \\ \hline
224  & 31  & 3  & 204  & 0.77 & 45.7 \\
598  & 33  & 6  & 74.1 & 0.85 & 18.4 \\
1365 &     & 3  & 11.2 & 18.2 & 59.3 \\
2403 &     & 6  & 22.9 & 3.18 & 21.2 \\
3031 & 81  & 2  & 27.5 & 3.63 & 29.0 \\
3351 &     & 3  & 7.59 & 10.1 & 22.3 \\
3368 & 96  & 2  & 7.59 & 11.6 & 25.6 \\
4321 & 100 & 4  & 7.59 & 16.1 & 36.2 \\
4536 &     & 4  & 7.59 & 16.7 & 36.8 \\
4639 &     & 4  & 2.82 & 25.1 & 20.6 \\
5457 & 101 & 6  & 28.8 & 7.38 & 61.8 \\ \hline
\end{tabular}
\end{center}
\caption{The Hubble type $T$, face-on angular 25 B-mag arcsec$^{-2}$ 
isophotal diameters $d_{25 {\rm (ang)}}$, Cepheid distances, $R$, and actual 
linear diameters $D_{25 {\rm (true)}}$ corresponding to this isophote, for 
11 local spiral galaxies of Hubble types 2 to 6.}
\end{table}
\end{center}

Consider now the distribution of (natural) log linear diameters for this
local sample of Table 1, supplemented by the Milky Way. We will use log
linear diameters throughout this paper since these have
been shown to be well-modelled by a Gaussian
distribution (c.f. Paturel 1979; Lynden-Bell \et 1988). Calculating
the sample mean value, $< \log D>$, and dispersion, $\sigma$, of the
(natural) log linear diameter distribution for this local sample, we 
obtain

\[
< \log D >  =  3.39 \quad \quad  \sigma  =  0.41
\]

Fig. \ref{figure:local} shows the sample cumulative distribution of
log linear diameter, compared with the cumulative distribution function of a
Gaussian with mean and dispersion equal to the sample values. It can be seen
that the local data give a good fit to a Gaussian distribution, as borne out
by the significance level (or $p$ value) of the Kolmogorov Smirnov (KS)
statistic.

\subsection{Distant galaxy sample}

If the nearby galaxy sample is typical of the universal intrinsic
population of galaxies in the same range of Hubble types, we can use this 
information to obtain an estimate 
of the Hubble parameter.  We have used angular diameters of more than 
3827 spiral galaxies out to a redshift of over 20000 km s$-1$ (although 
the vast majority of galaxies in the sample have redshifts of 1000 - 
5000 km s$^-1$) taken from the RC3 catalogue.

All these galaxies have known redshifts, so the assignment of their 
true linear diameters depends only on the assumed value of the Hubble 
Parameter to give their distances, together with the assumption
that the Hubble flow is uniform.  The sample of galaxies is spread across 
the sky including both cluster and field galaxies and we assume 
no bulk local motion that may change distances.  Unlike Sandage, we 
have included spirals of all Hubble types 2 through 6.  The justification 
for this is that the linear diameters for a chosen value of $H_0$ are  
the same for all of these Hubble types (de Jong 1996).  

\subsubsection{Completeness}

The sample of distant galaxies is biased strongly toward large galaxies as 
the RC3 catalogue has an angular diameter cut-off at $\approx 1$ arcmin. 
At a redshift distance
of 5000 km s$^{-1}$ this would correspond to a galaxy with diameter 
$14.5h^{-1}$ kpc.  
This result shows that galaxies even the size of the Milky Way will not 
appear in 1 arcmin angular diameter limited catalogues further away 
than $9200h$ km s$^{-1}$.  At even larger distances only the most 
superlarge galaxies (similar to NGC 1365 or M 101) would appear in 
such catalogues. We address the question of completeness in two different
ways, which we now summarise. A comprehensive discussion of our statistical
analysis will follow shortly in a full report.
\begin{enumerate}
\item{Under the assumption that the spatial distribution of galaxies is
uniform and the intrinsic distribution of log linear diameters is Gaussian,
we determine the probability density function of log linear diameter for 
{\em observable\/} galaxies subject to a selection function characterised by 
a sharp apparent angular diameter limit at 1 arcmin,
{\em and\/} an upper and lower redshift cutoff. The primary function of the
redshift cutoff was to exclude significant deviations from uniform
Hubble flow due to galaxies at low redshift with substantial peculiar
motions. As discussed by e.g. Gould (1993), the presence of a redshift
(or equivalently distance) selection function also introduces a bias in
the distribution of observable objects. Our analysis sets out to both
compute and remove this bias in conjunction with the more familiar `Malmquist'
bias directly associated with the angular diameter limit. We thus estimate
from the distribution of observable galaxies the parameters of the intrinsic,
Gaussian, distribution of log linear diameters.}
\item{As an alternative to the above approach, instead of attempting to
remove the selection bias to which the distant sample is subject, we 
subject the distant sample to a new selection limit, dependent
only on the log linear diameter. Since the selection is now only a function
of an intrinsic quantity we can apply the same selection criterion to the
local sample.}
\end{enumerate}

We discuss quantitative aspects of these two treatments of selection effects
in the Results section below.

\section{Results}

The extreme range of possible values for the Hubble parameter 
determined by different techniques and published recently is from 
about 40 to 80 km s$^{-1}$ Mpc$^{-1}$.  If the value 
of $H_0$ were as low as 40 km s$^{-1}$ Mpc$^{-1}$ then the Galaxy (and 
most other galaxies from the local calibrating sample) would not appear 
above the angular diameter cut-off at a redshift distance of
5000 km s$^{-1}$ Mpc$^{-1}$ or greater.
This would seriously undermine several other
recent determinations of $H_0$, as our local sample comprises e.g. those
galaxies used to calibrate the I band Tully-Fisher relation
(c.f. Giovanelli, 1996). If, however, 
$H_0$ were higher than 80 km s$^{-1}$ Mpc$^{-1}$, then the Milky Way and M31 
(which seem to be typical representatives of nearby galaxies over the
considered range of Hubble types, in the 
centre of our calibrating sample) would be large spirals and NGC 1365 
and M 101 would be among the largest spirals in the observable Universe.  

We have used three methods in concert to examine the size distribution 
of the distant galaxies and estimate $H_0$, quantifying the above
two general conclusions.  First a subset of distant 
galaxies between 1500 and 5000 km s$^{-1}$ was selected.  As mentioned in the
preceding section, the lower limit of  
this sample was chosen to eliminate the worst of the contamination of the 
sample by peculiar motions, whilst the upper limit was chosen so as to 
provide an approximately distance-limited sample. We found that the extension
of the upper limit to 40000 km s$^{-1}$ (i.e. effectively no upper redshift
limit!) yielded an observable distribution of log linear diameters which
deviated significantly from the Gaussian distribution, with a shifted
mean, predicted by our model of uniform spatial density and a sharp angular
diameter limit. We considered this to be due to the breakdown of our model
assumptions; in particular the selection function at high redshift was
unlikely to be depend only on angular diameter, but also on inclination
and would probably begin to display a sensitivity to morphological type.
When we adopted the limit of 5000 km s$^{-1}$, however -- while this
introduced a new selection bias similar to that discussed by Gould (1993)
and thus rendered
the predicted distribution of log linear diameters formally non-Gaussian --
we found that the computed observable distribution with this new selection
function showed deviations from Gaussianity of only a few percent. We thus 
continued to model the predicted observable distribution as a Gaussian.
Moreover, the distant sample now matched this predicted observable 
distribution very well. Figure \ref{figure:l04} 
shows the sample cumulative distribution of log linear diameter for the
selected galaxies and, for comparison, the best-fit Gaussian to the
predicted distribution of observable galaxies with this new selection
function. Note that application of the KS test suggests a good fit of the
data to this approximately Gaussian model. 

The observed mean of the best-fit Gaussian
distribution lies at $<\log D> = 3.48 \pm 0.01$, if we assume that 
$H_0 = 60$ km s$^{-1}$ Mpc$^{-1}$. Correcting for the effects of
bias introduced by the angular diameter and redshift selection limits,
we deduce that the best-fit mean value of the {\em intrinsic\/} distribution
of log linear diameter is $< \log D> = 3.25 \pm 0.02$. The mean of the 
calibrating sample, 
however, lies at $<\log D> = 3.39 \pm 0.12$, where the quoted error is
the standard error on the mean derived from the sample dispersion.
These two samples would, therefore, have the same mean if 

\[
H_0 = 52 \pm 6 {\rm km s}^{-1} {\rm Mpc}^{-1}
\]

As a further means of placing limits on the likely value of $H_0$ it is
useful to consider the {\em order statistics\/} of the modelled intrinsic 
distribution of log linear diameters (c.f. Hendry, O'Dell \& Collier-Cameron
1993). For a sample of 12 calibrating galaxies, which we are assuming to be
drawn from the same intrinsic population as the distant sample, one can
therefore pose the question of how likely it is, for a given value of $H_0$,
that one would obtain e.g. as small a galaxy as the galaxy with smallest
linear diameter, or conversely as large a galaxy as that with largest
linear diameter. There are many different ways to combine the order statistics
and apply statistical tests to their values, and we will describe such an
analysis in detail in our follow-up paper. Here we present one plot, which
succinctly illustrates the same principal features as a more sophisticated
analysis. Figure \ref{figure:order} shows the lower tail of the third order 
statistic (ie. the probability that the third smallest galaxy in a sample of 
size 12 drawn from the distant sample would be as small as, or smaller than,
the smallest observed member of the local calibrating sample) and the upper
tail of the tenth order statistic (ie. the probability that the third largest
galaxy in a sample of size 12 drawn from the distant sample would be as large 
as, or larger than, the largest observed member of the local calibrating
sample) for different assumed values of $H_0$ in the range 30 - 100 km 
s$^{-1}$ Mpc$^{-1}$. We could, of course, choose any of the order statistics
for this illustration but the third and tenth are suitable examples.
As can be seen from this figure, the value of $H_0$ where these two
probability curves cross is found at $H_0 \approx 48$ km s$^{-1}$ Mpc$^{-1}$.
In addition it can be seen that values of 
$H_0 > 75$ km s$^{-1}$ Mpc$^{-1}$ are not favoured (as the largest galaxies
in the local sample would then be unreasonably large compared with the
distant galaxies) nor are values of 
$H_0 < 35$ km s$^{-1}$ Mpc$^{-1}$ (as then the smallest galaxies in the
local sample would then be unreasonably small compared with the distant
galaxies). 
 
The above analysis assumes that we have correctly accounted for, and
corrected for, the observational selection effects within the distant sample,
and that the local sample is a ``fair'' sample of the intrinsic distribution
of log linear diameters. A further statistical analysis 
of the data was made by, instead, imposing a sharp linear diameter limit
on both the distant and nearby samples. We adopt a limit of $14.5h^{-1}$ kpc
since at the redshift limit of 5000 km s$^{-1}$) this corresponds to an
angular size of 1 arcmin -- above which we assume that the entire sample
is reasonably complete. Provided again that the local galaxies are
assumed to be
a fair sample of the intrinsic population, the distribution of log linear
diameter in the {\em truncated\/} local and distant samples should be
identical. This analysis is less powerful than that using the entire sample 
(especially at low $H_0$ where the number of calibrators above the 
cut-off diameter becomes small), but has the advantage of applying the 
same selection function to both samples instead of essentially
reconstructing the `missing' galaxies from the observed distribution. Thus,
what is lost in statistical power is partially gained in robustness, and we
consider this approach a useful consistency check.

Fig. \ref{figure:limit} shows a comparison of the means of the truncated
distant and local samples, together with $1-\sigma$ errors on the means, as
a function of $H_0$ (which, of course, determines the truncation limit). 
As can be seen from this figure, good agreement between the truncated means
of the local and distant data are found at $H_0 \approx 53$ and
$H_0 \approx 65 - 72$ km s$^{-1}$ Mpc$^{-1}$. This weaker analysis
therefore suggests that $50 < H_0 < 75$ gives a reasonable bound.
As above, however, one can place some further constraints on $H_0$ by
considering the order statistics of the largest galaxies in the local
sample, and on this basis we find that the lower end of this range is
favoured by the data. Assuming the local calibrators to be a fair sample of
the intrinsic
distribution, we carried out a series of Monte Carlo simulations in order to
determine the probability of a truncated random sample of 12 galaxies
larger than $14.5h^{-1}$ kpc in diameter containing two galaxies larger than
59 kpc in diameter, for a range of different values of $H_0$. This probability
is less than 10\% for all $H_0 > 65$ km s$^{-1}$ Mpc$^{-1}$ and drops below
5\% for $H_0 > 75$ km s$^{-1}$ Mpc$^{-1}$.

\section{Conclusions}

The results presented in this report strongly suggest that values of
$H_0$ higher than 75 km s$^{-1}$ Mpc$^{-1}$ or  
below 40 km s$^{-1}$ Mpc$^{-1}$ are ruled out. Clearly this is not a 
dramatically new conclusion, but should serve to underline that the
simple idea of comparing the linear size of local and distant galaxies
is certainly a plausible method for estimating $H_0$.
A value of $H_0$ in the range 50 - 55 km s$^{-1}$ Mpc$^{-1}$
gives a good fit between the mean log diameter of the
local and distant samples -- both with and without a lower diameter limit --
and renders the statistical properties of the largest and smallest local
galaxies easily compatible with those predicted from the distant sample.
This result is also consistent with a number of independent recent estimates
from other distance indicators.

Our analysis has, however, used distance moduli to the local galaxies
derived via the the traditional Cepheid calibration of Madore \& Freedman
(1991), adopting a true distance modulus of 18.5 for the Large Magellanic
Clouds. Recently the LMC distance modulus has been revised,
using Hipparcos data (Reid 1997; Gratton \et 1997) to calibrate the
Cepheid period-luminosity relation in galactic open clusters.  The simplest 
interpretation of the new results is that our estimate for $H_0$ should be 
reduced by about 10 per cent.  Taking this into account would imply 
that the true value of the Hubble parameter lies in the range 
45 - 50 km s$^{-1}$ Mpc$^{-1}$.

On the other hand, Kochanek (1997) has embarked on a comprehensive
re-analysis of the Cepheid distance scale and has concluded that
relative distance moduli between the LMC and a number of local galaxies,
as determined from HST Cepheid observations, have been significantly
{\em under\/} estimated. The effect of the Kochanek re-calibration of
the Cepheid distance scale may well largely cancel out any reduction in the 
Hubble parameter resulting from the Hipparcos results. For the moment, then,
we will consider the range $H_0 = 50 - 55$ km s$^{-1}$ Mpc$^{-1}$ as the
range most likely from this analysis.
In a simple Einstein-de Sitter universe with $\Omega = 1$ and 
$\Lambda = 0$ this range corresponds to an age of
approximately 12 - 13 Gyr since the Big Bang. 

For the value of 
the Hubble parameter to lie significantly outside of this range, 
not only the Milky Way but most of the spirals in our local
sample must be collectively either significantly larger or significantly 
smaller than the population of spirals in the visible Universe. 
If that were the case, it would mean that we live in an atypical part of the 
Universe, and that would remove the justification for any extrapolation from 
the nearby region to the Universe at large. Given that, on the contrary, we
find the Milky Way to be a very average-sized spiral galaxy compared with
other local galaxies of similar Hubble type, it seems unlikely to us that
the `principle of terrestrial mediocrity' should hold on the scale of
the local supercluster and yet break down on larger scales.

\section{Acknowledgements}

Electronic RC3 data was taken from the CDS internet site and we acknowledge 
the use of the Starlink computers at Sussex and Glasgow.  SPG would also 
like a postdoc, please.

\section{Bibliography}

\begin{trivlist}

\item  Freedman\,W.\,L. 1996, to appear in Critical Dialoges in Cosmology, ed.
N. Turok, LANL preprint no. astro-ph/9612024

\item Giovanelli\,R. 1996, to appear in The Extragalactiv Distance Scale,
eds. M.\,Livio, M.\,Donahue \& N.\,Panagia (Cambridge), LANL preprint no.
astro-ph/9610116

\item Goodwin\,S.\,P., Gribbin\,J., Hendry\,M.\,A. 1997, MNRAS, submitted, 
LANL preprint no. astro-ph/9704216

\item Gould\,A   1993, ApJ, 412, 55

\item Graham\,J.\,A., \et 1997, ApJ, 477, 535

\item Gratton\,R.\,G., Fusi Pecci\,F., Carretta\,C., Clementini\,G., 
Corsi\,C.\,E., Lattanzi\,M.  1997, preprint, LANL preprint no. 
astro-ph/9704150 

\item Hendry\,M.\.A., O'Dell\,M.\.A., Collier-Cameron\,A.    1993,
MNRAS, 265, 983

\item de Jong\,R.\,S.  1996, A\&A, 313, 45

\item Kochanek\,C.\.S.   1997, preprint, LANL preprint no.
astro-ph/9703059

\item Lynden-Bell\,D., Faber\,S.\.M., Burstein\,D., Davies\,R.,
Dressler\,A., Terlevich\,R.\.J., Wegner\,G.     1988, ApJ, 32, 19

\item Madore\,B\,F., Freedman\,W.\,L.   1991, PASP, 103, 933

\item Paturel\,G.    1979, A\&A, 74, 206

\item Reid\,N.  1997, preprint, LANL preprint no.
astro-ph/9704078

\item Sandage\,A.   1993a, ApJ, 402, 3

\item Sandage\,A.   1993b, AJ, 404, 419

\item Sandage\,A.   1996, AJ, 111

\item de Vaucouleurs\,G., de Vaucouleurs\,A., Corwin\,H.\,G., Buta\,R.\,J., 
Paturel\,G., Fouque\,P.  1991, Third Reference Cataloge of Bright 
Galaxies (RC3).

\item van der Kruit\,P.\,C. 1987, in The Galaxy, eds.  G.\,Gilmore \& 
B.\,Carswell (NATO), p 27

\item van der Kruit\,P.\,C. 1990, in The Galactic and Extragalactic
Background Radiation, eds. S.\,Bowyer and C.\,Leinert (IAU) p 85

\item Vilenkin\,A. 1995, Phys Rev Lett, 74, 846

\end{trivlist}

\newpage

\begin{figure}
\caption{the fit of a gaussian of $<\log D> = 3.39$ and $\sigma = 0.41$ 
to the sample of local calibrating galaxies.}
\label{figure:local}
\end{figure}

\begin{figure}
\caption{Comparison of the sample cumulative distribution function of
log linear diameter, and the cumulative distribution of the predicted
best-fit Gaussian model, for 1388 
galaxies between 1500 and 5000 km s$^{-1}$. The Gaussian has a
mean of 3.48 and $\sigma = 0.42$, assuming $H_0 = 60$ km s$^{-1}$ Mpc$^{-1}$.}
\label{figure:l04}
\end{figure}

\begin{figure}
\caption{The change in the lower tail probability of the third order 
statistic (dashed line) 
and the upper tail probability of the tenth order statistic
(full line) with $H_0$.}
\label{figure:order}
\end{figure}

\begin{figure}
\caption{The means, enclosed by 1 $\sigma$ error bands, of the distant
(full line) and local (dashed line) galaxies when the sample is truncated at
a linear diameter of $14.5h^{-1}$ kpc, for $30 < H_0 < 100$.}
\label{figure:limit}
\end{figure}

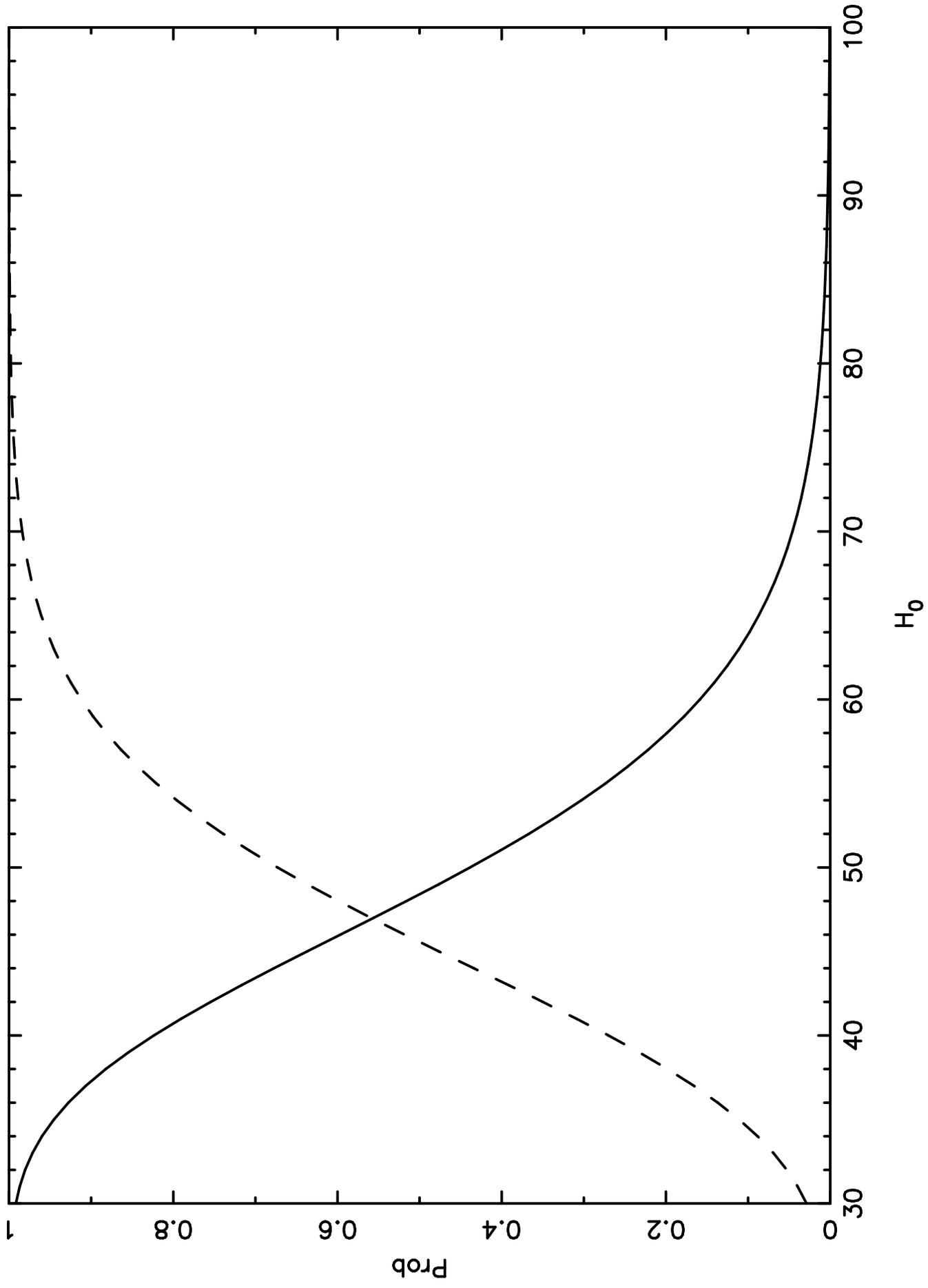
\end{document}